\documentclass[useAMS,usenatbib,dcolumn,usegraphicx,twocolumn,10pt]{mnarxiv}
\usepackage{float}
\usepackage{times}
\usepackage{amsmath}
\usepackage{placeins}
\usepackage{float,xspace}
\usepackage{hyperref}
\newcommand{\ud}[1]{\mathrm{d}#1}

\newcommand{\br}[1]{\left(#1\right)}

\renewcommand{\sq}[1]{\left[#1\right]}
\newcommand{\msun}{\,\mathrm{M}_{\sun}}
\newcommand{\gu}{\,\mathrm{km}\,\mathrm{s}^{-1}\kpc^{-1}}
\newcommand{\gradu}{\gu}

\newcommand{\K}{\textbf{\textsf{K}}}
\newcommand{\E}{\textbf{\textsf{E}}}

\newcommand{\kpc}{\,\mathrm{kpc}}

\newcommand{\pd}[2]{\frac{\partial}{\partial{#2}}{#1}}

\newcommand{\hi}{H{\small I}\xspace}
\newcommand{\hhi}{H{\scriptsize I}\xspace}
\renewcommand{\exp}[1]{\mathrm{e}^{#1}}
\renewcommand{\vec}[1]{\bmath{#1}}

\title[]{The role of large-scale magnetic fields in galaxy NGC 891: can magnetic
fields help to reduce the local mass-to-light ratio in the galactic outskirts?}
\author[]{Joanna Ja{\l}ocha$^{1}$,
{{\L}ukasz Bratek$^{1}$\thanks{E-mail:
Lukasz.Bratek@ifj.edu.pl}},
{Jan P\c{e}kala$^{1}$},
{Marek Kutschera$^{2}$}
\\
$^{1}$The H. Niewodnicza\'{n}ski Institute of Nuclear Physics,
Polish Academy of Sciences, Radzikowskego 152, PL-31342 Krak\'{o}w, Poland\\
$^{2}$The M. Smoluchowski Institute of
Physics, Jagellonian University,  Reymonta 4, PL-30059 Krak{\'o}w, Poland}

\begin{document}
\date{\today}
\pagerange{\pageref{firstpage}--\pageref{lastpage}} \pubyear{2009}

\maketitle

\begin{abstract}
We address the problem of the influence of large-scale magnetic fields on galactic rotation for the example of the spiral galaxy NGC 891. Based on its rotation curve and the surface density of \hi we determine, in the framework of the global disc model, the surface density of matter. Then, based on the surface brightness, we determine the corresponding profile of the local mass-to-light ratio. We also model the vertical gradient of azimuthal velocity in the quasi-circular-orbit approximation, and compare it with measurements. We discuss what factors may influence the rotation of matter in NGC 891 and how this can translate to changes in the profile of the local mass-to-light ratio. In particular, we discuss the possible effect of magnetic fields on the motion of ionized gas, and, consequently, on the determination of the profile of the local mass-to-light ratio. Finally, we put forward the hypothesis that the asymmetry in magnetic fields observed in NGC 891 might be responsible for the observed anomalous behaviour of the vertical gradient.
\medskip
\hrule
\begin{small}
\flushleft \textbf{The definitive version is available at\\ \url{http://onlinelibrary.wiley.com/doi/10.1111/j.1365-2966.2012.20447.x/abstract}}
\end{small}
\medskip
\hrule
\end{abstract}


\section{Introduction}

NGC 891 is a nearby edge-on spiral galaxy. Its very high inclination
angle allows the vertical gradient of the azimuthal component
of the velocity field to be determined. Measurements include
gradients of $-15\gradu$ \citep{bib:Oosterloo}, $-18.8 \gradu$ \citep{bib:grad891} and
$-17.5\pm5.9\gradu$ for $z\in\br{1.2,4.8}\kpc$, $r\in\br{4.02,7.03}\kpc$ in the NE quadrant
\citep{bib:Benjamin}.

In \citep{bib:gradient} we modelled the vertical gradient for the galaxy NGC 891 for the global disc model and quasi-circular-orbit approximation, assuming that the motion of gas above the disc plane is governed mainly by the gravitation of the column mass density distributed in the mid-plane. Our gradient estimate was in agreement with the above measurements. This coincidence led us to suggest that NGC 891 was a flat, disc-like object, devoid of a massive spheroidal component, at least in the region covered by the gradient measurements (outside the central bulge). This qualitative result concerning the mass distribution is consistent with a more recent finding suggesting that in NGC 891 there is no dark matter within $r=15\kpc$ \citep{bib:fraternali}.

The first aim of this paper is to see to what extent our previous gradient prediction is altered by new measurements. To reconstruct the column mass density we use the rotation curve determined in \citep{bib:fraternali} and the previous measurement of neutral hydrogen taken from \citep{bib:Oosterloo}. In section \ref{sec:gradient} we show that with the new data our gradient estimate is not significantly altered.

Our approach to mass modelling is different from the usual one presented in \citep{bib:fraternali}. We do not assume a constant mass-to-light ratio and we do not derive the mass distribution based on brightness measurements. First, we obtain the mass distribution directly from the rotation measurements and the distribution of gas. Next, based on the surface brightness measurements, we determine \textit{local} mass-to-light ratios, which then can be drawn as a function of galactocentric radial distance.

It is a feature of the global disc model that the profile of local mass-to-light ratios is sensitive to the accuracy in determining the rotation curve. A small variation in the rotational velocity can lead to a significant reduction in the mass-to-light ratio in the galactic outskirts. On the other hand, it is known \citep{bib:nature,bib:battaner} that magnetic fields can strongly influence the rotational field of the gas. More importantly, as concluded in \citep{bib:nature}, the flatness of rotation curves for the outer parts of discs of spiral galaxies could be the result of interstellar magnetic fields. As we will see, the inclusion of magnetic fields can help to reduce the usually high value of the local mass-to-light ratio at the disc 'rim'.

Regarding galaxy NGC 891, there is some peculiarity in the behaviour of the vertical gradient of rotation. The gradient vanishes in the SW arm. We test our hypothesis that this vanishing could be caused by the presence of a relatively intense magnetic field of a dozen or so microgauss.

The final conclusion of this paper is that our results are consistent with the predictions of \cite{bib:fraternali} that there is no dark matter within $15\kpc$. Furthermore, non-baryonic dark matter could be absent also at larger radii, as high local mass-to-light ratios (often associated with the presence of dark matter) can be reduced to reasonably small values merely by taking into account the interaction of the gas with magnetic fields. We come to this conclusion by applying a simplified model. It seems that the non-trivial task of rigorously taking into account galactic magnetic fields might turn out to be important for our understanding of galactic dynamics.

\section{Modeling of the vertical gradient of rotation}\label{sec:gradient}

\begin{figure}
\vspace{-0.53cm}
\hspace{-0.6cm}
\includegraphics[width=0.54\textwidth]{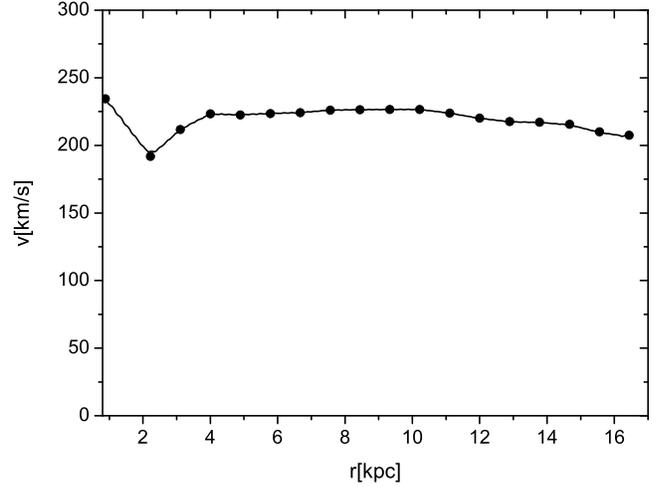}
\vspace{-0.9cm}
\caption{\label{fig:rot}
Rotation curve of spiral galaxy NGC~891. The solid circles denote
the measurement points taken from \citep{bib:fraternali}, and the line is
the model rotation curve obtained in the global disc model.} \end{figure}

\begin{figure} \vspace{-.53cm} \hspace{-0.6cm} \includegraphics[width=0.53\textwidth]{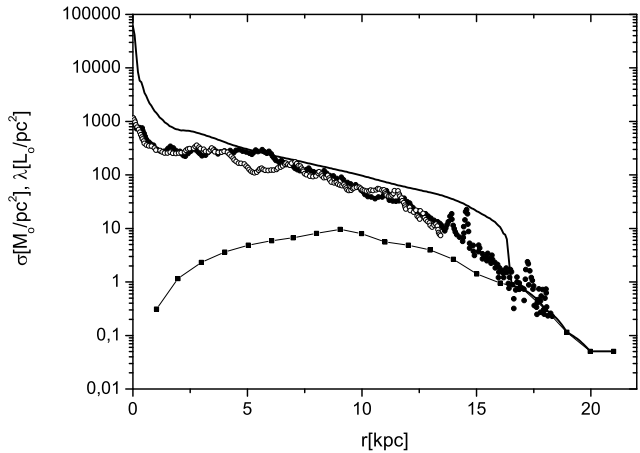} \vspace{-1.0cm} \caption{\label{fig:sig} Surface mass density in the global disc model calculated based on the rotation curve taken from \citep{bib:fraternali} \textit{[solid line]}; the surface brightness in the $3.6\mathrm{\mu{}m}$ filter taken from \citep{bib:fraternali} shown separately for the NE quadrant \textit{[full circles]} and for the SW quadrant \textit{[open circles]}; the surface density of \hhi+He based on \citep{bib:Oosterloo} \textit{[joined full squares]}. } \end{figure}

We reconstruct the substitute surface mass density in the global disc model based on the rotation curve \citep{bib:fraternali} shown in Fig.\ref{fig:rot}, and the measured surface mass density of neutral hydrogen \citep{bib:Oosterloo} multiplied by a factor of 4/3 to account for the presence of  helium. The relationship between the rotation curve $v_c(r)$ and the surface mass density $\sigma(r)$ in disk model, where $r$ is the galactocentric distance in the disk plane, is given by the following integral\footnote{$\E$ and $\K$ are complete elliptic integrals of the first and second kind: $$ {\K}(k)=\int\limits_{0}^{\pi/2}\frac{\ud{\phi}}{\sqrt{1-k^2 \sin^2{\phi}}},\qquad {\E}(k)=\int\limits_{0}^{\pi/2}\sqrt{1-k^2 \sin^2{\phi}}\ \ud{\phi}.$$} \citep{bib:bratek_MNRAS}
\begin{equation} \label{eq:SigmafromRotCrv} \!\! \sigma(r)= \frac{1}{\pi^2G} \lim\limits_{\epsilon\to0}\left[\int\limits_0^{r-\epsilon} v_{c}^2(\chi)\biggl(\frac{\K\br{\frac{\chi}{r}}}{ r\ \chi}-\frac{r}{\chi} \frac{\E\br{\frac{\chi}{r}}}{r^2-\chi^2}\biggr)\ud{\chi}+\dots \right.\end{equation}
\vspace{-0.03\textheight} \begin{equation*} \phantom{XXXXXXXXX}\left. \dots + \int\limits_{r+\epsilon}^{\infty}v_{c}^2(\chi) \frac{\E\br{\frac{r}{\chi}} }{\chi^2-r^2}\,\ud{\chi}\right].\end{equation*}
Here, $\chi$ is an integration variable expressed in the same units as $r$. It is important to remember that when the rotation data are cut off, so is the integration region, the result of integration contains a boundary term that can differ from that in another integral with the derivative of $v_{c}^2$ and given, for example, in \citep{bib:binney}. The two integrals are equivalent, provided they are not cut off. After the integration has been formally cut off, they give different results because of the singularities in the integral kernels. However, the difference is not important for radii out to $60$ per cent of the cutoff radius \citep{bib:bratek_MNRAS}. This shows that the disc model must be used with due care, as the result depends on what is assumed about the rotation curve beyond the cutoff radius. The integration should not be merely cut off. Moreover, when the integration is cut off then the calculated mass density can be lower than observed for hydrogen in the vicinity of the cutoff radius. To minimize the uncertainty, we find the surface density iteratively, using a method similar to that described in \citep{bib:jalocha_apj}. The method finds a mass density that agrees with the distribution of gas for radii greater than the cutoff radius, and the rotation calculated for this extended mass distribution agrees with the rotation observed for radii smaller than the cutoff radius. The results are shown in Fig.\ref{fig:sig} together with the surface density of neutral hydrogen multiplied by a factor of $4/3$, separately for the NE and SW quadrants.

\begin{figure} \vspace{-.38cm} \hspace{-0.6cm} \includegraphics[width=0.53\textwidth]{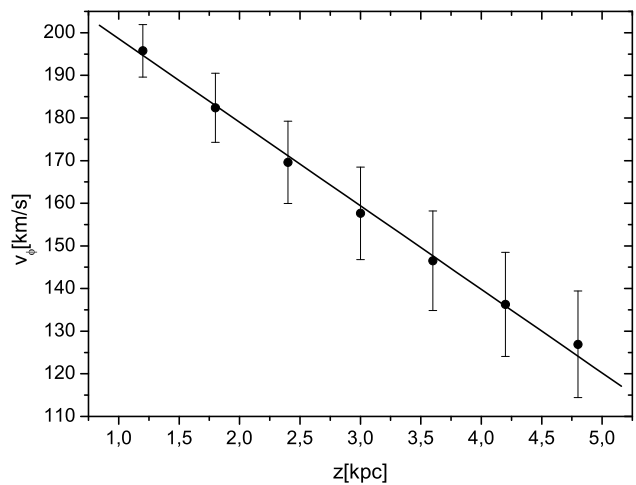} \vspace{-0.8cm} \caption{\label{fig:grad}The azimuthal component of rotation as a function of the vertical distance from the mid-plane. The points represent the velocity values averaged over the interval $r\in(4.02, 7.03)\kpc$, and the bars show a standard deviation in that interval.} \end{figure}

In the quasi-circular-orbit approximation the rotational velocity close to the mid-plane is given by by
\begin{equation}\label{eq:VOverdiscFromSigma} {v_{\varphi}^2(r,z)}=\!\int\limits_0^{\infty}\!\!\!\frac{2\, G\sigma\br{\chi{}}\chi{}\ud{\chi{}}}{\sqrt{\br{r+\chi{}}^2+z^2}}\!\br{\!\K(k)-\! \frac{\chi^2-r^2+z^2}{\br{r-\chi}^2+z^2}\,\E(k)\!}\!,\phantom{XXXXXX} \end{equation}
\vspace{-0.03\textheight} \begin{equation*} 0\leq{k}=\sqrt{\frac{4r\chi{}}{\br{r+\chi{}}^2+z^2}}<1.\end{equation*}
This relation allows us to determine the vertical gradient of the azimuthal velocity, $\partial_zv_{\varphi}$. In practice it can be calculated in two ways, similarly to in \citep{bib:gradient}. In method I we calculate $v_{\varphi}$ for an array of points $(r_i,z_j)$. Then for each $z_j$ we find an average over all $r_i$ in the specified range of $r$. The results are shown in Fig.\ref{fig:grad}. The error bars represent a standard deviation. With the use of a linear fit the gradient estimate is found to be $-19.6\pm2.9\gu$. In method II, gradients are found separately for every $r_i$ with the use of a linear fit. Next, the mean value of the gradients and the uncertainty defined as one standard deviation are calculated. In that case the gradient estimate is $-19.16\pm1.75\gu$.
For comparison, we recall the previous result with another rotation curve: $-19.9\pm3\gu$ and $-19.5\pm1.7\gu$, respectively \citep{bib:gradient}. The result is only slightly changed, which is to be expected as the two rotation curves are similar in the region of interest. This change is insignificant within uncertainties.

It is worthwhile to stress that it is a feature of disc models that they give high gradient values. Moreover, our result shows that the azimuthal velocity is a linear function of $z$, and the relatively small standard deviation in method II shows that the gradient is weakly dependent on the radial variable. These characteristics are consistent with measurements.

\section{The profile of local mass-to-light ratio}

\begin{figure}\vspace{-.58cm} \hspace{-0.6cm} \includegraphics[width=0.54\textwidth]{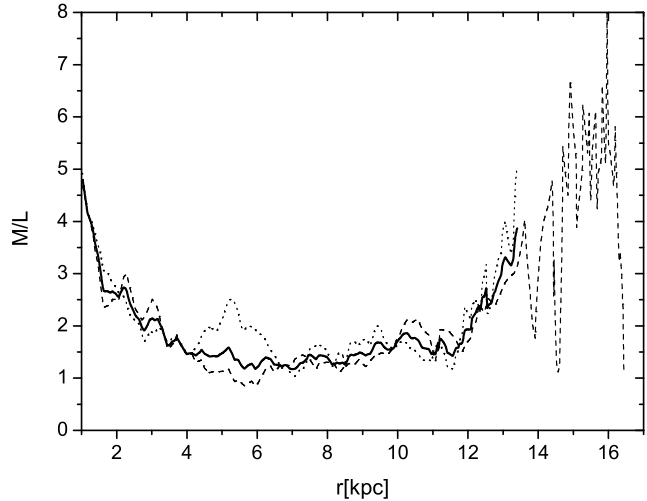} \vspace{-0.8cm} \caption{\label{fig:mtol} The profile of the local mass-to-light ratio (without \hhi+He). NE quadrant  \textit{[dashed line]}; SW quadrant \textit{[dotted line]}; the mean value \textit{[solid line]}.} \end{figure}

The profile of the local mass-to-light ratio for NGC 891 is shown in Fig.\ref{fig:mtol}, separately for the NE and SW arms (together with the averaged profile).
To obtain it we used the $3.6\mathrm{\mu{}m}$ surface brightness profile of NGC~891 published in \citep{bib:fraternali}, corrected for the inclination of $88^{\circ}$ (LEDA), and the surface mass density found in the previous section after subtracting the contribution from neutral hydrogen and helium (non-luminous in the $3.6\mathrm{\mu m}$ filter). The local mass-to-light ratio is $5$ in the central region (outside $1\kpc$ where the rotation and hydrogen data are available), then it falls off rapidly to $1-2$ for $r\in(3,12)\kpc$ and finally rises to a value above $4$ at $r=13.4\kpc$. For large radii the surface brightness is known only for the NE quadrant. The brightness measurement is not very accurate and shows a large spread, making the local mass-to-light ratio oscillate rapidly between $1$ and $8$.

What conclusions can be drawn? First, the high inclination angle almost certainly causes the surface brightness profile to be underestimated (extinction processes can be decisive). Furthermore, molecular hydrogen and other gases are present and not accounted for -- if they were, the local mass-to-light ratio would decrease even more. Taking this into account, and also that the ratio is low in most parts of the galaxy, we see that there is no need to introduce non-baryonic dark matter. The rapid increase in the local light-to-mass ratio profile in the outermost region of the rotation curve is also not worrying -- for large radii in the NE quadrant the average local mass-to-light ratio is above $5$, which is not greater than in the central region.

\begin{figure}\vspace{-.53cm} \hspace{-0.65cm} \includegraphics[width=0.54\textwidth]{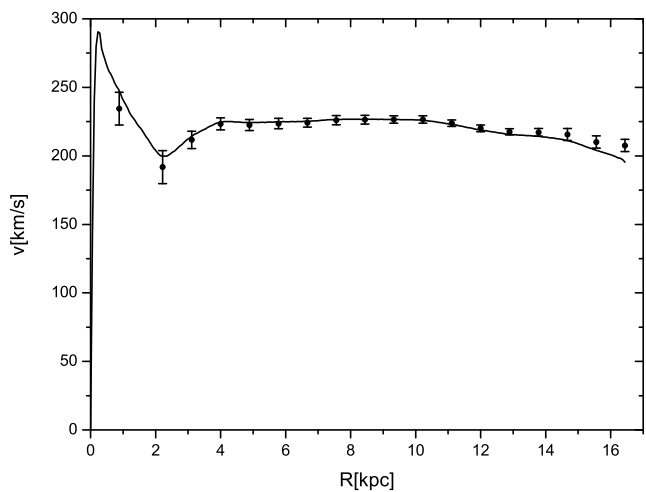} \vspace{-0.8cm}\caption{\label{fig:rotzmod} A small  modification of rotation velocities \textit{[solid line]}. Points with error bars were taken from \citep{bib:fraternali}.} \end{figure}

\begin{figure}\vspace{-.43cm} \hspace{-0.65cm}  \includegraphics[width=0.54\textwidth]{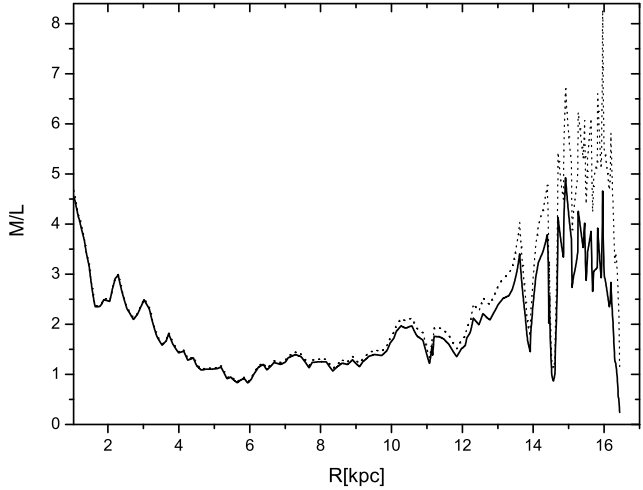} \vspace{-0.8cm}\caption{\label{fig:mtolmod} A change in the profile of the local mass-to-light ratio caused by a small modification of the rotation \textit{[solid line]}, compared with the original profile of the local mass-to-light ratio \textit{[dotted line]}. Here, only the NE quadrant is considered.} \end{figure}

We stress that for large radii the local mass-to-light ratio is sensitive even to very small variations in the rotation curve. To show this, we change the rotation curve slightly by decreasing it in the outer regions, but not by much, so that, except for a single point, the profile agrees with measurements within error limits (see Fig.\ref{fig:rotzmod}). Even such a small change can significantly decrease the profile of local the mass-to-light ratio (see Fig.\ref{fig:mtolmod}).

\subsection{The role of magnetic fields}

As we will see below, despite the presence of ionized gas above the galactic plane, the influence of magnetic fields with intensities of a few microgauss can be ignored in the central region where the motion of gas is dominated by gravity. In particular, such small magnetic fields should not play any role in the vertical gradient of rotation for NGC 891 in the range $r\in\br{4.02,7.03}\kpc$. Therefore, we expect our gradient model to work well in this region, accounting satisfactorily for gradient measurements. At larger radii, however, the situation is different and including the magnetic fields can be essential.

We have seen that the profile of the local mass-to-light ratio in the galactic outskirts is strongly dependent on variations in the rotation curve. However, the rotation curve in this region is determined based on measurements of gaseous mass components (mainly  hydrogen).
 If the gas is at least partially ionized, its motion will be influenced by magnetic fields \citep{bib:nature,1995MNRAS.277.1129B}. This effect is especially prominent at large radii \citep{bib:battaner}. For a galaxy with a typical mass of about $10^{11}\msun$, the influence of magnetic fields even at a distance of $10\kpc$ can be significant compared with gravitational forces \citep{bib:kutschera}. In what follows we will estimate whether taking this magnetic effect into account can affect the local mass-to-light ratio determination.

\begin{figure}\vspace{-.53cm} \hspace{-0.65cm}  \includegraphics[width=0.54\textwidth]{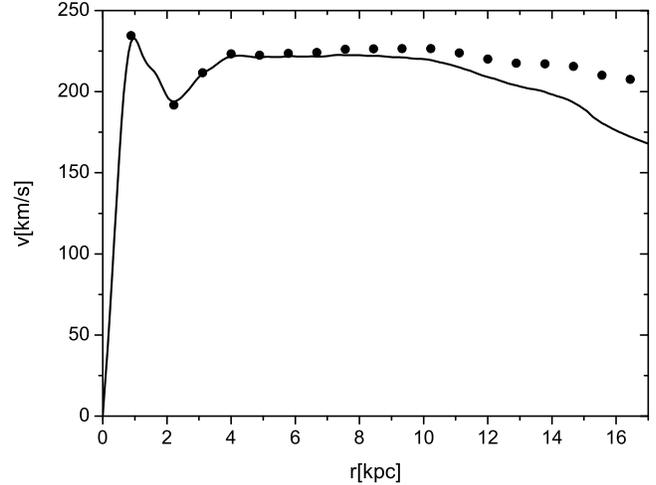} \vspace{-0.8cm}\caption{\label{fig:pole1} The measured rotation curve \citep{bib:fraternali} \textit{[circles]} and rotation velocities lowered so that the local mass-to-light ratio will be a non-increasing function of galactocentric distance \textit{[solid line]} } \end{figure}

\begin{figure}\vspace{-.57cm} \hspace{-0.65cm}  \includegraphics[width=0.54\textwidth]{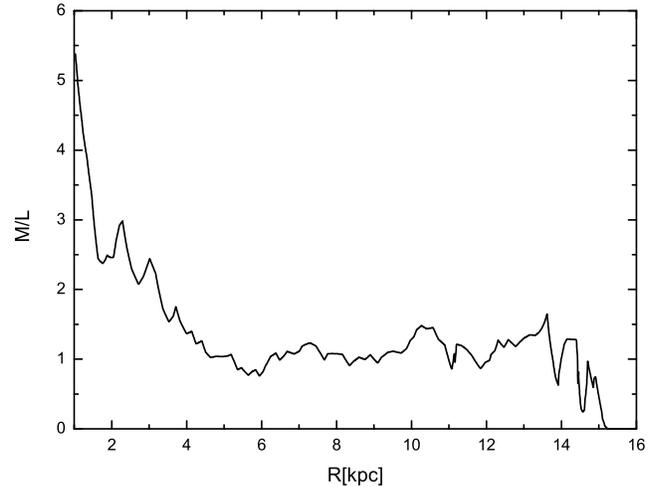} \vspace{-0.8cm}\caption{\label{fig:pole2} A non-increasing profile of the local mass-to-light ratio corresponding to a rotation curve suitably changed for large radii as shown in Fig.\ref{fig:pole1}.} \end{figure}

\begin{figure}\vspace{-.53cm} \hspace{-0.65cm} \includegraphics[width=0.54\textwidth]{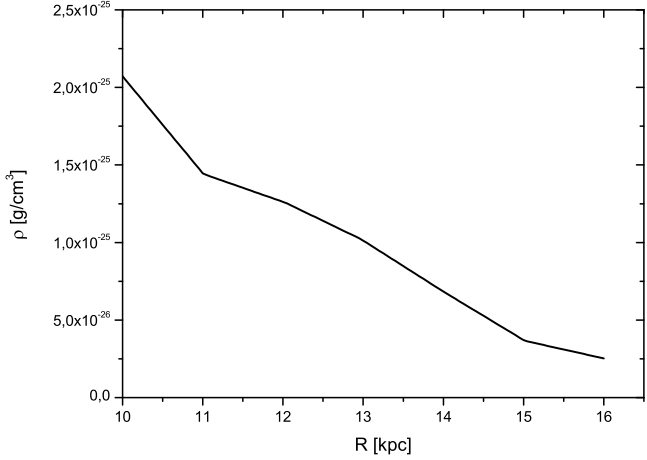} \vspace{-0.8cm}\caption{\label{fig:rozwodoru} Bulk density of \hhi in the disk plane obtained based on the surface density given in \citep{bib:Oosterloo}. } \end{figure}

\begin{figure}\vspace{-.53cm} \hspace{-0.65cm}  \includegraphics[width=0.54\textwidth]{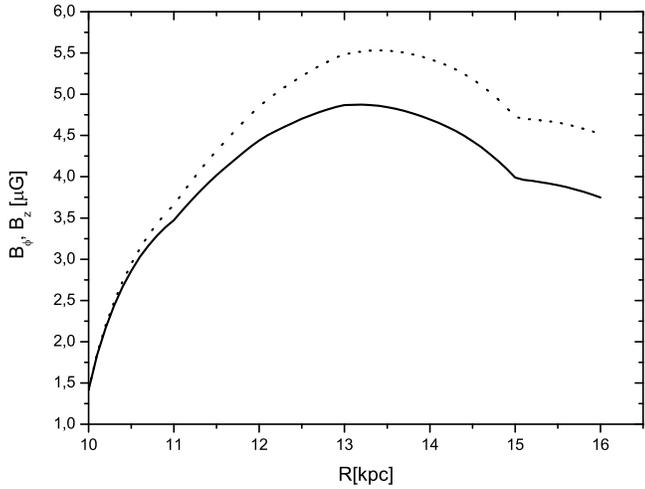} \vspace{-0.8cm}\caption{\label{fig:graph1} Magnetic field needed to lower the rotation so that the profile of the local mass-to-light ratio is a non-increasing function of the galactocentric distance: the vertical component $B_{z}$ of magnetic field  [dotted line], and the azimuthal component $B_{\phi}$ of the magnetic field [solid line].} \end{figure}

Consider an example rotation curve that agrees with the rotation curve of NGC 891 everywhere except at larger radii, where we assume the rotation to be slower than observed (see Fig.\ref{fig:pole1}).
The resulting surface mass density is such that the predicted profile of the local mass-to-light ratio is a non-increasing function of radius (see Fig.\ref{fig:pole2}),
in contrast to the situation shown in Fig.\ref{fig:mtol} with the original rotation curve. Now, let us calculate how strong the magnetic field should be to cause such a change in the rotation curve. The starting point is the stationary pressure-less and viscous-less Navier–Stokes equation: \begin{equation}\label{eq:navier} (\vec{v}\circ\vec{\nabla})\vec{v}=-\vec{\nabla}\Phi +\frac{1}{4\pi\rho}{\br{\vec{\nabla}\times\vec{B}}\times\vec{B}}. \end{equation} Here, $\rho$ is the density of gas. It is assumed that the large-scale magnetic field $\vec{B}$ is given. Close to the disc, assuming the velocity field to be of the form $v_{\varphi}(r)\vec{e}_{\varphi}$, and the magnetic field to be of the form $B_{\varphi}(r)\vec{e}_{\varphi}+B_{z}(r)\vec{e}_{z}$, the radial component of these equations gives \[-\frac{V_{\varphi}^2(r)}{r}=-\pd{\Phi}{r}-\frac{1}{8\pi\rho}\sq{\frac{1}{r^2}\pd{\br{r^2 B_{\varphi}^2(r)}}{r} +{\pd{B_{z}^2(r)}{r}}},\] and the other components are identically satisfied. Introducing the notation $V_{\varphi}^2=v_{\varphi}^2+\br{\delta{}v_{\varphi}}^2$ and denoting by $v_{\varphi}^2$ the contribution from the gravitational field, we have \[\frac{\br{\delta{}v_{\varphi}(r)}^2}{r}=\frac{1}{8\pi\rho}\sq{\frac{1}{ r^2}{\pd{\br{r^2 B_{\varphi}^2(r)}}{r}} +\pd{B_{z}^2(r)}{r}}.\] Here $(\delta v_{\varphi})^2$ is the contribution to the rotational velocity from the magnetic field. First, consider the case of an azimuthal field ($B_z=0$). Then the required magnetic field to account for $(\delta v_{\varphi})^2$ would be \begin{equation}\label{eq:navier2}B_{\varphi}(r) = \frac{1}{r} \sqrt{(r_{1})^2 (B_{\varphi}(r_{1}))^2 + 8 \pi   \int\limits_{r_1}^r \rho(\xi)(\delta v_{\varphi}(\xi))^2  \xi d \xi}. \end{equation} We assumed an exponential falloff of the bulk density $\rho$ along the direction normal to the disc plane: $\rho=\rho_0\exp{-z/h}$, with $h=1\kpc$. Then $\rho_0\approx\sigma/{2h}$, where $\sigma$ is the surface density of hydrogen \hi measured  in \citep{bib:Oosterloo} (see Fig.\ref{fig:rozwodoru}).
The result of the numerical integration of equation \eqref{eq:navier2} is shown in Fig.\ref{fig:graph1}.

It can be seen that a magnetic field of the order of a few microgauss suffices to produce the required difference in velocities. A similar calculation can be repeated assuming a purely vertical magnetic field with $B_z$ being a function of the radius only. In that case, equation (\ref{eq:navier}) simplifies to \begin{equation}\label{eq:navier3}(\delta v_{\varphi})^2=\frac{r}{8\pi\rho} \pd{B_{z}^2 }{r} \end{equation} with the solution \begin{equation}\label{eq:navier4}B_{z}(r) = \sqrt{ (B_{z}(r_{1}))^2 + 8 \pi \int\limits_{r_1}^r \rho(\xi)(\delta v_{\varphi}(\xi))^2 \frac{1}{\xi} d \xi}. \end{equation} {In equation \eqref{eq:navier4} we assume the same density $\rho$  as in equation \eqref{eq:navier2}.} The result of the numerical integration is presented in Fig.\ref{fig:graph1}. Similarly to in the previous case, the field is of the order of several microgauss. In the galaxy NGC 891 there is indeed a large-scale magnetic field present of the order of a few microgauss (both in the disc and above it) \citep{bib:pole,bib:pole1,bib:pole2}.

With the magnetic field as shown in Fig.\ref{fig:graph1}, the mass-to-light ratio profile remains constant, but weaker fields can reduce this ratio. The plots in Fig.\ref{fig:graph1} are shown for illustrative purposes only. They allow estimation of the strength of the magnetic field required to reduce the mass-to-light ratio.

The situation considered above serves as a toy model, illustrating the fact that the presence of magnetic fields with an intensity of a few microgauss cannot be simply ignored, as they have a significant impact on the motion of (even only partially) ionized gas. Not accounting for this effect may lead to an overestimation of the profile of the local mass-to-light ratio.

\section{The vertical gradient of rotation and magnetic field}\label{sec:gradientsecondary}

We recently applied our gradient estimation method to six galaxies with available measurements of the vertical gradient of rotation \citep{bib:gradient,2011MNRAS.412..331J}. We found our predictions of the gradient value to agree within error bars with measurements. This agreement was reached without taking into account the influence of magnetic fields. This is so because the gradient was measured only at small distances from the galactic centres (eg. $4\!\!-\!\!7\kpc$ for NGC~891 or $3\!\!-\!\!8\kpc$ for Milky Way). As will become evident later, and as follows from a simple estimation made in \citep{bib:kutschera}, magnetic fields with a typical intensity of several microgauss are too weak, compared with the gravitational influence, to have a noticeable effect. For gradient measurements at low altitudes above the disc plane (as for the Milky Way), the influence of magnetic fields can be neglected with greater certainty when the gas density decreases with the altitude above the galactic plane.  In Section \ref{sec:gradient} we had such an agreement with gradient measurements for galaxy NGC 891 in the NE quadrant.

\begin{figure} \vspace{-0.53cm} \hspace{-0.65cm}   \includegraphics[width=0.54\textwidth]{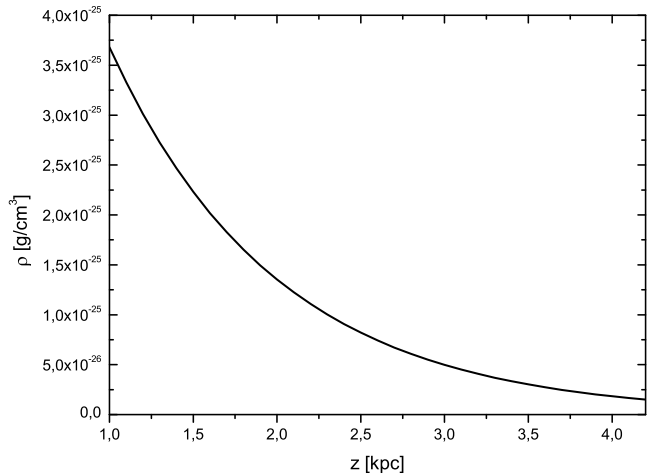} \vspace{-0.65cm}\caption{\label{fig:gradnewro}  The bulk density of \hi as a function of vertical distance from the disk plane.} \end{figure}

\begin{figure}\vspace{-.53cm} \hspace{-0.65cm}  \includegraphics[width=0.54\textwidth]{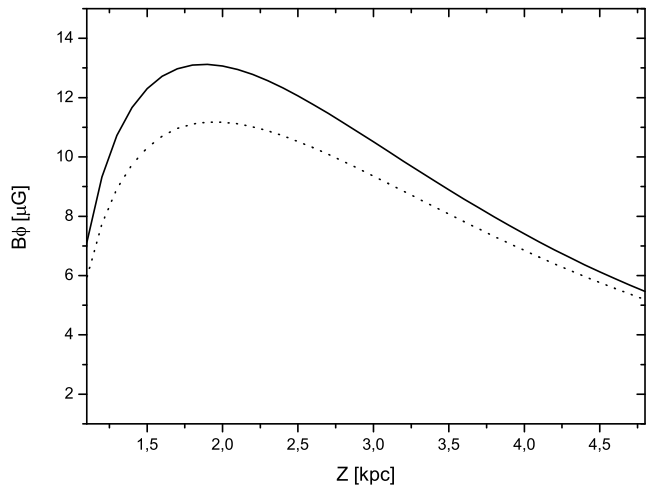} \vspace{-0.8cm}\caption{\label{fig:polegrad}  Magnetic field required to cancel the vertical gradient shown at $r=4.0\kpc$ [solid line] and at $r=7.0\kpc$  [dotted line]. {Calculations of the field assume the gas density as in Fig.\ref{fig:gradnewro}.} } \end{figure}

There can be stronger magnetic fields present in galaxies, of the order of a dozen or so microgauss \citep{bib:battaner}. Such strong fields can be decisive for the motion of ionized gas, in particular of gas of low density. A gradient is not observed in the SW arm of NGC 891. Could a magnetic field be responsible for this anomaly, reducing the gradient to zero? To answer this question we again employ the Navier–Stokes equation \eqref{eq:navier}. For simplicity, we can assume a purely azimuthal magnetic field that depends on the variable $z$ only. Then: $$\tilde{v}^2_{\phi}(r,z)=-r\pd{\Phi(r,z)}{r}+\frac{B_{\phi}^2(z)}{4\pi\rho},\quad -r\pd{\Phi(r,z)}{r}={v}^2_{\phi}(r,z),$$ where $\tilde{v}^2_{\phi}(r,z)$ is the rotation velocity observed above the disc, and ${v}^2_{\phi}(r,z)$ is the rotation velocity one would expect in the absence of a magnetic field. A constant gradient value is observed above the mid-plane, starting from $z_a=1\kpc$. Hence,  $\tilde{v}^2_{\phi}(r,z)=\tilde{v}^2_{\phi}(r,z_a)$ for $z>z_a$. This gives us \begin{equation}\label{eq:givesus} B_{\phi}^2=B_{\phi}^2(z_a)+4\pi\rho\br{{v}^2_{\phi}(r,z_a)-{v}^2_{\phi}(r,z)}.\end{equation} The density $\rho$ in this equation is shown in Fig.\ref{fig:gradnewro}.
We assumed an exponential falloff of the density along the direction normal to the disc plane, with $h=1\kpc$. In fact, the obtained $B_{\phi}$ is also a function of $r$; however, it is weakly dependent on $r$ compared with the changes associated with the variable $z$. Hence, the result for $B_{\phi}$ can still be regarded as consistent with the earlier assumption that $\partial_rB_{\phi}=0$. The behaviour of $B_{\phi}$ is shown in Fig.\ref{fig:polegrad}.

It can be seen that to reduce the gradient of rotation in the region $r\in\br{4,7}\kpc$, a magnetic field with the intensity of a dozen or so microgauss would suffice. This means that we need fields larger than typical for spiral galaxies. The results of \cite{bib:pole} and \cite{bib:pole2} suggest that particularly strong and ordered fields might indeed be locally present in the SW arm of galaxy NGC 891. Similarly to in the previous section, we stress that the field values are shown for the purpose of illustration only. They show the order of magnitude of the magnetic field required to cancel the gradient. With this in mind, we can conclude that a field of a dozen or so microgauss can affect the gradient and, if there were such a field in the SW arm of NGC 891, it could be used as a possible explanation for the anomalous gradient. There is indeed a certain field asymmetry observed: 'in the south-western part of NGC 891 the polarization percentage is between 20 and 30 per cent, implying very well ordered B-field' \citep{bib:pole2}. If the asymmetry applies also to the field intensity, then we would have a strong premiss for attributing the gradient asymmetry to this magnetic effect.

\section{Summary}

From the above it follows that galaxy NGC 891 is a disc-like object without a dominant massive non-baryonic dark matter halo. Two facts support this statement. First, the vertical gradient values of the azimuthal velocity calculated in the framework of the disc model are high, consistent with the measurements. Second, the obtained profile of the local mass-to-light ratio is low. Although the ratio rises for larger radii, its behaviour, as we have shown, is sensitive to small variations in rotational velocities. To illustrate this, we reduced the rotation by a small amount (allowed by measurement errors). Moreover, the ratio profile may be overestimated by the omission of factors such as the presence of gases other than \hi or the influence of magnetic fields.

From the above it follows that galaxy NGC 891 is a disc-like object without a dominant massive non-baryonic dark matter halo. Two facts support this statement. First, the vertical gradient values of the azimuthal velocity calculated in the framework of the disc model are high, consistent with the measurements. Second, the obtained profile of the local mass-to-light ratio is low. Although the ratio rises for larger radii, its behaviour, as we have shown, is sensitive to small variations in rotational velocities. To illustrate this, we reduced the rotation by a small amount (allowed by measurement errors). Moreover, the ratio profile may be overestimated by the omission of factors such as the presence of gases other than \hi or the influence of magnetic fields.

Our gradient modelling accounts well for measurements in the NE quadrant. Regarding the SW quadrant, the absence of gradient could be associated with the presence of stronger magnetic fields, with intensities of a dozen or so microgauss. If such asymmetry occurs for a galaxy, one can hypothesize that it is caused by some asymmetry in the intensity or configuration of the large-scale magnetic field. Based on the research in \citep{bib:pole2,bib:pole} it seems there is such an asymmetry in NGC 891. We stress that our considerations concerning magnetic fields are illustrative in character. The influence of magnetic fields on the motion of matter depends on the field intensity, its configuration, and on the degree of gas ionization and the gas density. On the one hand, accurate determination of these factors is not possible; on the other hand, the influence of magnetic fields on galactic rotation cannot be ignored.

In galaxy NGC 891 not only are 'there are no visible effects of dark matter out to 15 kpc' \citep{bib:fraternali}, but most probably there is also no massive non-baryonic dark matter halo, although we cannot rule out its presence completely.

\bibliography{magnetic}
\bibliographystyle{mn2e}

\end{document}